\documentclass[useAMS,usenatbib]{mn2e}
\usepackage{longtable,color}
\usepackage{graphicx}
\usepackage{mathrsfs}
\usepackage{epsfig}
\usepackage{natbib}
\usepackage{color}
\usepackage{float}
\usepackage{amsmath}
\usepackage{times}
\usepackage{upgreek}
\usepackage[varg]{txfonts}
\usepackage{enumerate}
\usepackage{verbatim}
\usepackage{booktabs}
\usepackage{multirow}
\usepackage{amssymb}
\usepackage{placeins}

\title[QG at $z\sim 3.7$]{Massive Quiescent Galaxies at $z>3$ in The Millennium Simulation Populated by A Semi-analytic Galaxy Formation Model}
\author[Rong et al.]{Yu Rong$^{1}$\thanks{E-mail: rongyu@bao.ac.cn}, Yingjie Jing$^{1}$, Liang Gao$^{1,2}$, Qi Guo$^{1}$, Jie Wang$^{1}$, Shuangpeng Sun$^{1}$, \and Lin Wang$^{1}$, Jun Pan$^{1}$\\
$^{1}$Key Laboratory for Computational Astrophysics, National Astronomical Observatories, Chinese Academy of Sciences, Beijing 100012, China\\
$^{2}$Institute of Computational Cosmology, Department of Physics, University of Durham, Science Laboratories, South Road, Durham DH1 3LE, UK
}

\begin{document}
\maketitle

\begin{abstract}
We take advantage of the statistical power of the large-volume dark-matter-only Millennium simulation, combined with a sophisticated semi-analytic galaxy formation model, to explore whether the recently reported $z=3.7$ quiescent galaxy ZF-COSMOS-20115 (ZF; Glazebrook et al. 2017) can be accommodated in current galaxy formation models. In our model, a population of quiescent galaxies (QGs) with stellar masses and star formation rates comparable to those of ZF naturally emerges at redshifts $z<4$. There are two and five ZF analogues at the redshift $3.86$ and $3.58$ in the Millennium simulation volume, respectively. We demonstrate that, while the $z>3.5$ massive QGs are rare (about 2\% of the galaxies with the similar stellar masses), the existing AGN feedback model implemented in the semi-analytic galaxy formation model can successfully explain the formation of the high-redshift QGs as it does on their lower redshift counterparts. 

\end{abstract}
\begin{keywords}
methods: numerical \--– galaxies: evolution \-- galaxies: high-redshift \-- galaxies: star formation
\end{keywords}
\section{Introduction}

At high redshifts, most of the observed galaxies have enhanced star formation, interactions, and active galactic nucleus (AGN) activities (e,g, Elbaz et al. 2007; Tran et al. 2010; Gr\"utzbauch et al. 2011; Koyama et al. 2013; Martini et al. 2013; Bufanda et al. 2017). The redshift $z\sim 2$ marks an important epoch of galaxy evolution, encompassing both the peak of cosmic star formation rate (e.g., Madau et al. 1996, 1998; Hopkins \& Beacom 2006) and AGN activity (e.g., Croom et al. 2004; Barger \& Cowie 2005; Fanidakis et al. 2012). The cosmic star formation rate (SFR) is observed to drop sharply after $z=2$, which is expected to be caused by exhaustion of gas reservoir \citep{Menci08} or by supernovae or AGN feedback \citep{Neistein06,Dekel06,Menci06,vandeVoort11}. 

An observed population of quiescent galaxies (QGs) with suppressed star formation beyond $z\simeq 2$  has drawn much interest \citep[e.g.,][]{Labbe05,Daddi05,Kriek08a,Kriek08b,Kriek09,Fontana09,Straatman14,Nayyeri14,Bell15,Kriek16,Strazzullo16,Kado-Fong17}. In particular, \cite{Glazebrook17} recently reported the spectroscopic confirmation of a high-redshift massive QG ZF-COSMOS-20115 (hereafter ZF) at $z=3.717$ with a stellar mass $M_{\star}\sim 1.7\times 10^{11}\ M_{\odot}$, extremely low star formation rate SFR$<4\ M_{\odot}/\rm{yr}$, and short star formation timescale $<250$~Myr. When checking the current hydrodynamic simulations of galaxy formation, such as Illustris, EAGLE, MUFASA, MassiveBlack-II \citep{Wellons15,Park15,Dave16}, we find that none of these simulations contains such an object with a stellar mass and SFR comparable to those of ZF. However, these hydrodynamic simulations have relative small volumes ($\lesssim (100\rm{Mpc})^3$), which may not be sufficient to accommodate the ZF analogues. 

In this letter, we use the publicly available galaxy catalog of Guo et al. (2011) based on the Millennium simulation (MS; Springel et al. 2005), to explore whether the ZF analogues can be accommodated in the current $\Lambda$CDM galaxy formation model. The semi-analytic galaxy formation model of Guo et al. (2011) has been proven successful in reproducing many galaxy properties both in the local Universe and at high redshifts \citep[e.g.,][]{Guo11,Yates12,Guo13,Xie15,Buitrago17,Rong17}, and MS provides a simulation volume about 300 times larger than that of the Illustris, which provides a much better statistics to facilitate the comparison between the model and data.  

The letter is organized as follows. In section 2, we briefly introduce the simulation and galaxy formation model used in this study. In section 3, we present our results on searching for the ZF analogues at $z>3$ in our simulation, and investigate the formation of the high-redshift massive QGs. Our results will be summarized and discussed in Section 4. Throughout this letter, we use ``log'' to represent ``$\rm{log}_{10}$''.


\section{Simulation}

MS follows $2160^3$ dark matter particles with a particle mass resolution $8.6\times 10^8\ M_{\odot}/h$ from $z=127$ to 0 within a comoving box with a length of $500/h$~Mpc on a side. This volume is about $300$ times larger than that of Illustris. Particle data are stored at 64 logarithmically spaced output times. The cosmological parameters of MS adopt the first-year {\it {Wilkinson Microwave Anisotropy Probe}} (WMAP) results: $\Omega_{\rm{tot}}=1$, $\Omega_{\rm{m}}=0.25$, $\Omega_{\rm{b}}=0.045$, $\Omega_{\Lambda}=0.75$, $h=0.73$, $\sigma_8=0.9$, $n=1$. Although the cosmological parameters adopted in MS are slightly different from the recent results, this will not qualitatively change our main results (see the discussion in section~4).  

At each snapshot, dark matter halos are identified with a standard friend-of-friend (FOF) method with a linking length 0.2 times of the mean inter-particle separation (Davis et al. 1985). Then the SUBFIND algorithm is applied to identify the local overdense and self-bounded subhalos (Springel et al. 2001). Merger trees are constructed by linking each subhalo at a snapshot to its unique descendant using the algorithm described in Boylan-Kolchin et al. (2009). The semi-analytic galaxy formation model developed by Guo et al. (2011; Guo11) is then implemented on the merger trees to generate the galaxy catalogs. Here we briefly summarize the main physical processes relevant to quenching star formation in massive galaxies.

There is growing evidence that galactic nuclear activity is closely related to galaxy formation, and in particular, AGN feedback may effectively quench star formation in massive galaxies \citep[e.g.,][]{Smethurst16,Dubois13}. In Guo11, there are two BH growth modes: the `quasar' mode and `radio' mode. The quasar mode applies to BH growth in gas-rich mergers, during which the central BH of the major progenitor grows both by cannibalizing the BH of the minor progenitor and by accreting cold gas. The associated feedback due to this mode is only approximated as the energy input produced by starburst. The radio mode growth occurs through hot gas accretion onto a central BH, at a growth rate of $\dot{M}_{\rm{BH}}\propto M_{\rm{BH}}M_{\rm{hot}}$, where $M_{\rm{BH}}$ and $M_{\rm{hot}}$ are the masses of the central BH and hot gas, respectively. The power of radio mode feedback is assumed to be 10\% of the accreted energy $\dot{M}_{\rm{BH}}c^2$ ($c$ is the light speed). In Guo11, BH of a galaxy is initialized with a zero mass when the galaxy emerges, and the BH mass growth is not triggered until the galaxy undergoes a merger event.


\section{Results}
\subsection{The model massive quiescent galaxies at $z>3$}

\begin{figure}
\centering
\includegraphics[scale=0.45]{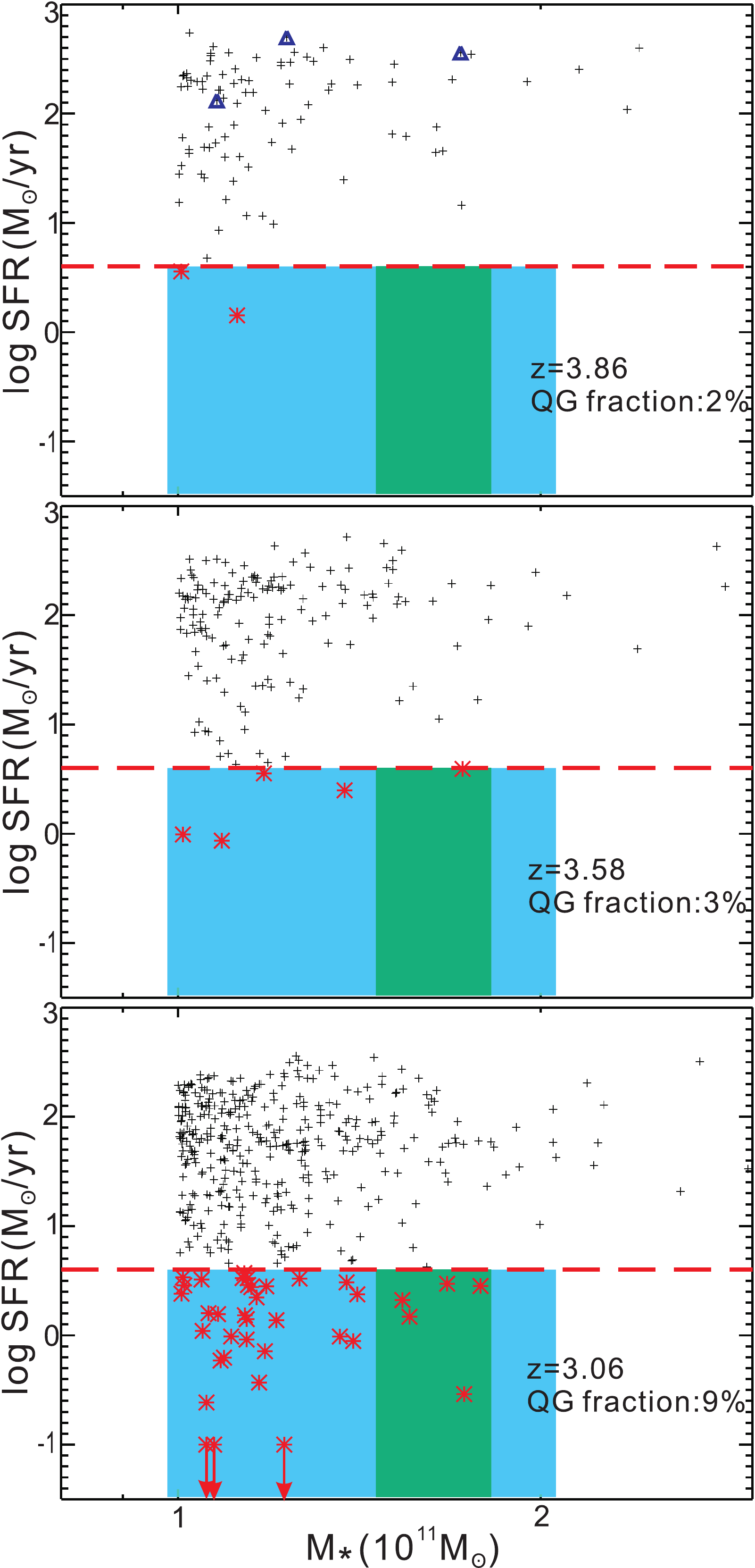}
\caption{SFR-$M_{\star}$ diagrams of the model massive galaxy sample ($M_{\star}>10^{11}\ M_{\odot}$) at $z=3.86$ (top panel), $3.58$ (median panel), and $3.06$ (bottom panel). The red stars and black crosses denote the galaxies with SFR$<4\ M_{\odot}/\rm{yr}$ and $\geq 4\ M_{\odot}/\rm{yr}$, respectively. The arrows indicate that the actual SFR values are out of the range of plot. The red horizontal dashed lines mark the upper-bound of the observed ZF, SFR$=4\ M_{\odot}/\rm{yr}$. The green and cyan shaded areas show the $1\sigma$ and $3\sigma$ stellar mass error ranges of ZF (Glazebrook et al. 2017), respectively. The blue triangles in the top panel show the three selected SFGs with their BH masses even higher than those of the ZF analogues at $z=3.86$.}
\label{sfr_stellarmass}
\end{figure}

From the model galaxy catalog, we first select a massive galaxy sample with stellar masses $M_{\star}>10^{11}M_{\odot}$ in the redshift range of $3<z<4$; $M_{\star}=10^{11}M_{\odot}$ is approximately at the lower threshold of $3\sigma$ stellar mass range of the observed ZF. In Fig.~\ref{sfr_stellarmass}, we show the stellar mass vs. SFR relations for the selected model massive galaxies at the three epochs: $z=3.86,3.58$, and $3.06$, as labeled in the different panels. The red horizontal dashed lines show the upper limit of SFR of ZF, $\sim 4\ M_{\odot}/\rm{yr}$, which is estimated with the Hydrogen Balmer H$\beta$ line flux. The green and cyan shaded areas bracket the $1\sigma$ and $3\sigma$ error ranges of the stellar mass of ZF, respectively. At $z=3.86$, there are $91$ galaxies more massive than $10^{11}\ M_{\odot}$ in the MS volume; 2 of them have SFR$<4\ M_{\odot}/\rm{yr}$, and they are slightly less massive than the $1\sigma$ stellar mass estimation of ZF, but still within the $3\sigma$ mass estimation bound. At $z=3.58$, there are $160$ massive galaxies, and 5 of them have SFR comparable to that of ZF, and one of them has a stellar mass within the $1\sigma$ stellar mass error range of ZF.

The seven $z\geq 3.58$ model galaxies with SFR$<4\ M_{\odot}/\rm{yr}$ and stellar masses $M_{\star}>10^{11}M_{\odot}$ are referred to as ZF analogues, and their properties are listed in Table~\ref{infor}. The fractions of the ZF analogues in our massive galaxy sample are as low as about $2\--3\%$ at the two redshifts, implying that the observed ZF is indeed quite rare. The number of the ZF analogues increases quite rapidly towards lower redshifts: at $z=3.06$, there are $410$ massive galaxies in the whole simulation volume, while $37$ (approximately 9\%) of them have SFR similar to that of ZF, suggesting a rapid formation of massive QGs during a very short timescale at the redshifts $z>3$.

\begin{table} \center \normalsize
\begin{tabular}{@{}ccccc@{}}
\hline
\hline
$z$ & Identifier & $M_{\star}\ (10^{11}\ M_{\odot})$ & SFR $M_{\odot}/\rm{yr}$\\
\hline
3.86 & Q1 & 1.01 & 3.59  \\ 
     & Q2 & 1.12 & 1.42  \\
\hline
     & Q1$'$ & 1.01 & 0.98 \\ 
     & Q2$'$ & 1.09 & 0.86 \\
3.58 & Q3$'$ & 1.18 & 3.57  \\ 
     & Q4$'$ & 1.38 & 2.50  \\	
	 & Q5$'$ & 1.72 & 3.92  \\	 
\hline
\hline
\end{tabular}
\caption{Properties of the model ZF analogues at $z=3.86$ and $3.58$.}
\label{infor}
\end{table}

\subsection{Formation of quiescent galaxies in $\Lambda$CDM}

\begin{figure}
\centering
\includegraphics[scale=0.4]{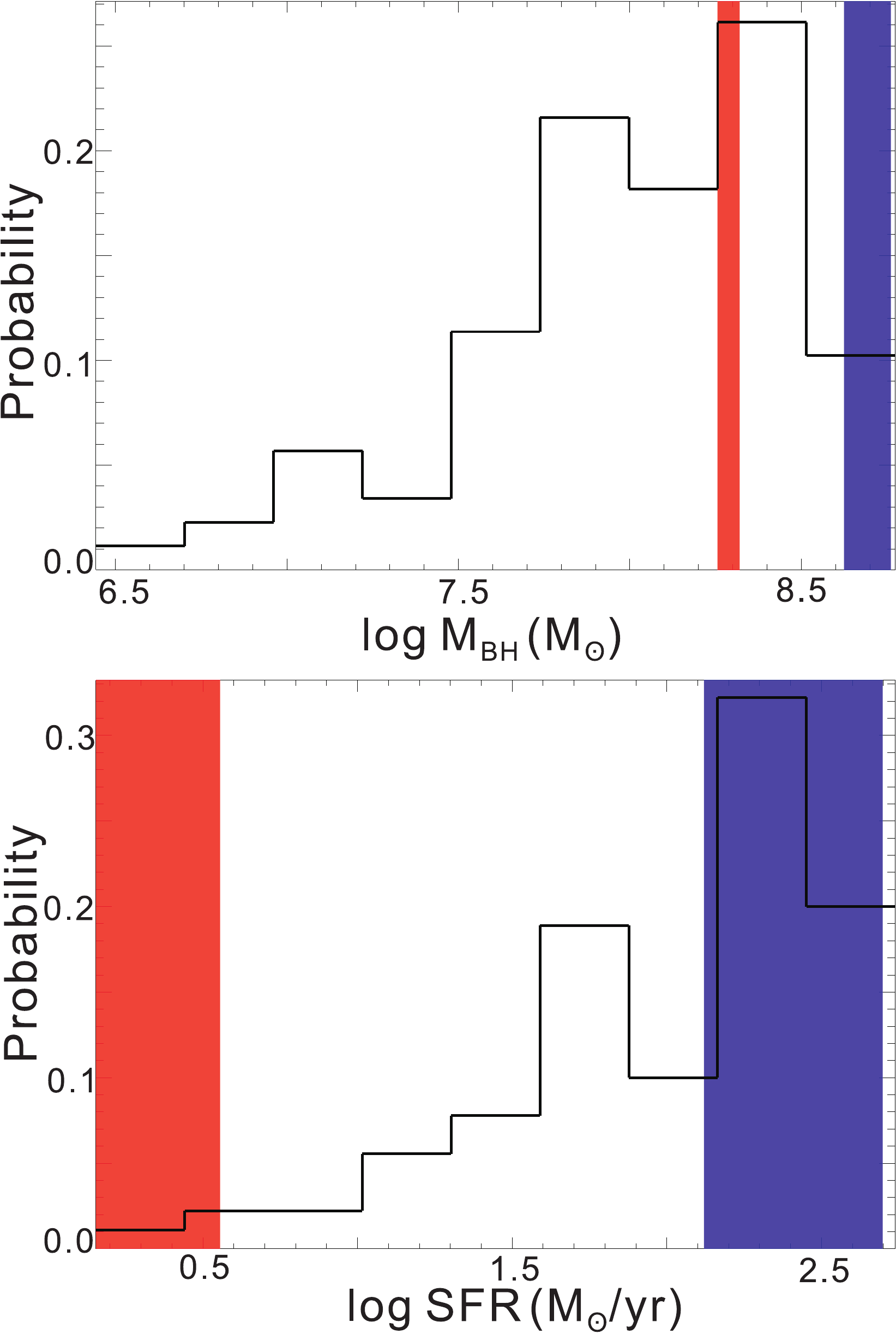}
\caption{The distributions of BH masses (upper) and SFR (lower) for the model galaxies with stellar masses more massive than $10^{11}\ M_{\odot}$ at $z=3.86$. The red and blue shaded areas show the ranges of the two ZF analogues and three SFGs with the most massive BHs, respectively.}
\label{property}
\end{figure}

A massive galaxy may contain a central supermassive BH, which can exert strong AGN feedback and remarkably affect the galaxy formation processes. It has been demonstrated that AGN feedback is crucial to understand many galaxy formation problems (Croton et al. 2006; Bower et al. 2008, Benson \& Bower 2011), e.g., AGN feedback can explain why star formation has been quenched in most of the present-day massive galaxies. Therefore, it is reasonable to expect that the model ZF analogues at high redshifts are also resulted from AGN feedback. Since the intensity of AGN feedback is assumed to be proportional to the mass of BH, $M_{\rm{BH}}$, in Fig.~\ref{property} we present the distributions of $M_{\rm{BH}}$ and SFR for the model massive galaxies at $z=3.86$. The BH masses and SFR of the high-redshift massive model galaxies cover a broad range, and the median value of $M_{\rm{BH}}$ is about $\sim 10^8 M_{\odot}$, consistent with that of the observed galaxies with the similar stellar masses \citep[e.g.,][]{Lauer07,Gultekin09,Sani11,Trakhtenbrot15}. For the two ZF analogues at $z=3.86$, their BH masses are about $10^{8.3}\ M_{\odot}$, relatively higher than the median value. We bracket the $M_{\rm{BH}}$ range of the two ZF analogues with the red shaded area in Fig.~\ref{property}. Interestingly, there are quite a few model galaxies having BH masses even higher than those of the ZF analogues, while being star-forming galaxies (SFGs) with SFR as high as $\gtrsim 100\ M_{\odot}/\rm{yr}$. We select the three SFGs with the highest BH masses (shown by the blue shaded areas in Fig.~\ref{property}) and compare their evolution histories with those of the two ZF analogues below. 

\begin{figure}
\centering
\includegraphics[scale=0.4]{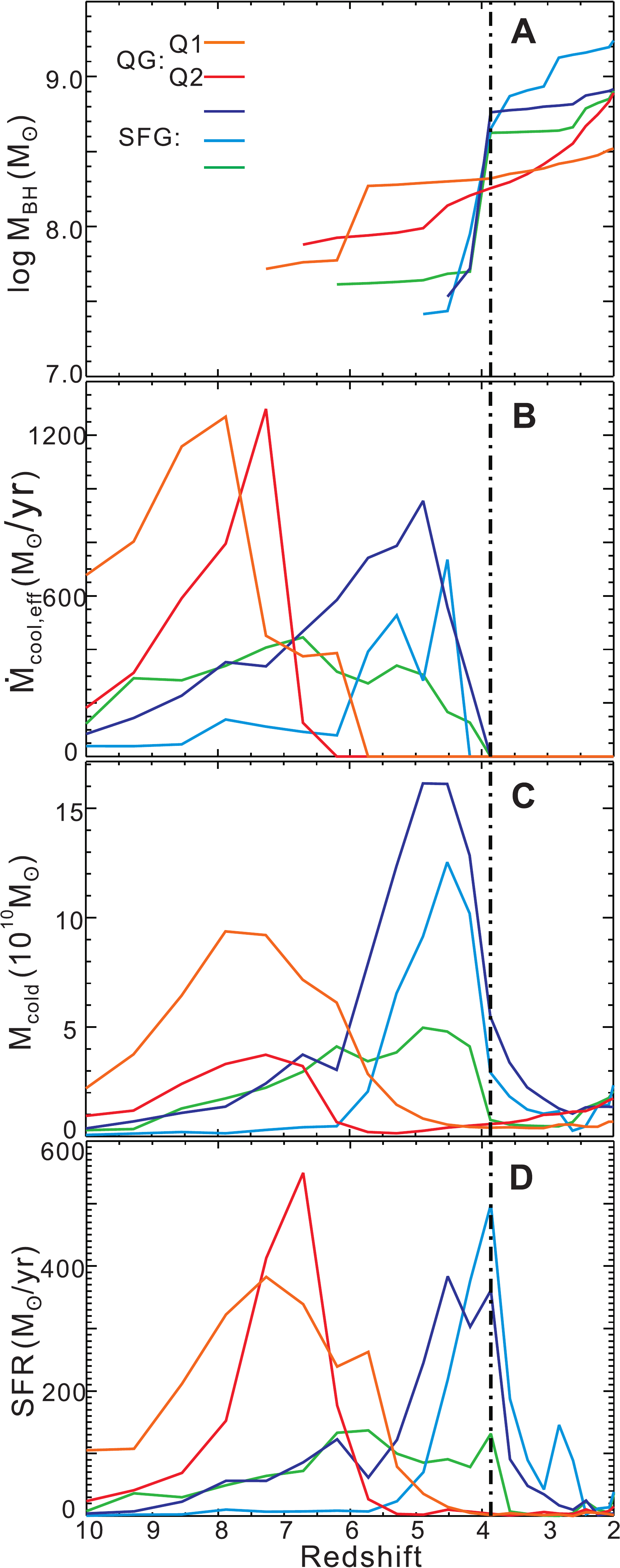}
\caption{Evolution of the five selected galaxies. A. black hole mass; B. Effective cooling rate; C. residual cold gas mass; D. SFR. The different galaxies are distinguished by the different colors.The dot-dashed line denotes $z=3.86$ at which these galaxies are selected.}
\label{evolution}
\end{figure}

In Fig.~\ref{evolution}, we compare the evolutionary properties of the five selected galaxies.
Panel A shows the growth of the central BH in each galaxy. Note that in many semi-analytic models, such as the model we used in this study, the growth of a central AGN is triggered by a galaxy merger event, i.e., the BH of a galaxy is initialized with a zero mass once the galaxy appears, and its mass growth is not triggered until a merger event occurs. 

Panel B shows the evolution of the effective cooling rate of gas in each galaxy, $\dot{M}_{\rm{cool,eff}}$, which is defined as $\dot{M}_{\rm{cool}}-\dot{M}_{\rm{AGNfeedback}}$, where $\dot{M}_{\rm{cool}}$ is the cooling rate of hot gas. When a central supermassive BH accretes the surrounding materials, it can deposit energy in a relativistic jet or bubble, and then the energy is delivered to heat the cooling gas at a rate of $\dot{M}_{\rm{AGNfeedback}}$, preventing any further gas fueling into the galaxy. Hence $\dot{M}_{\rm{cool,eff}}$ drops rapidly when AGN feedback kicks in and eventually reduces to 0.

Panel C shows the residual cold gas $M_{\rm{cold}}$ in each galaxy. In our model, the residual cold gas is not affected by AGN feedback and continues to form stars until it is exhausted. Panel D shows the evolution of SFR of each galaxy. The vertical dashed lines in all panels denote $z=3.86$. Compared with the three SFGs, the BHs in the two ZF analogues emerge at earlier epochs; therefore, the gas cooling is shut off earlier, and the systems have the required time to consume the residual cold gas and eventually become quiescent at $z\simeq 3.86$. For the three SFGs, since their BHs grow up at relatively later epochs, therefore though they have even more massive BHs by $z=3.86$, their residual cold gas has not been completely consumed. Consequently, they are still star-forming at $z=3.86$, yet they will become quiescent shortly after $z\simeq 3.86$, as shown in panel D.

In summary, in our model the formation of the $z=3.86$ quiescent galaxies requires relatively earlier merger events (at $z\sim 7$) in order to trigger earlier BH growth, so that AGN feedback could kick in earlier and exhaust the residual cold gas in these galaxies at $z = 3.86$.

\section{Discussion and summary}

In this letter, we make use of the statistical power provided by the large volume Millennium simulation, combined with a sophisticated semi-analytic galaxy formation model, to explore whether the recently reported $z=3.7$ quiescent galaxy ZF can be accommodated in the current galaxy formation models. 

In our model, a population of massive galaxies with the stellar masses comparable to that of ZF naturally emerges after $z=4$. At the redshift $z=3.86$, there are $91$ galaxies with the stellar masses similar to that of ZF in the simulation volume. Only $2$ ($2\%$) of them have SFR comparable to that of ZF, whilst the others are star-forming galaxies. The number of the ZF analogues in our model increases rapidly; by the redshift $3.58$ and $3.06$, there are $5$ and $37$ ZF analogues in the simulation volume, respectively. We demonstrate that our implementation of AGN feedback can successfully explain the formation of the high-redshift ZF analogues, as it does on the lower redshift massive QGs. We note that the different cosmological parameters have moderate effect on our conclusion. When using the model galaxy catalog of Henriques et al. (2015) which adopts the PLANCK Cosmology, the formation of ZF analogues is only slightly delayed in that model, with the $z=3.65$ ZF analogues population similar to that of our own ones at $z=3.86$.

The comoving volume of the Millennium simulation is $(500/h~\rm{Mpc})^3$, and there are 5 ZF analogues at $z=3.58$ in the entire simulation volume, corresponding to a space density of $1.6\times 10^{-8}~\rm{Mpc}^{-3}$. This value is much lower than that of the observed density claimed by \cite{Straatman14}, $\sim 1\times 10^{-6}~\rm{Mpc}^{-3}$. However, we argue below that the number density of the ZF analogues has large uncertainties both in the observation and theoretical model.

There are two uncertainties in determining the number density of the ZF analogues in observation. First, the survey of ZFOURGE \citep{Straatman14} only covers a small area of about $0.1^{\circ}$. As the quite rare objects, the clustering of the massive QGs is indeed strong at $z>3$, and therefore a much larger survey volume than ZFOURGE is needed to securely obtain the accurate observational number density. Second, SFR of the QGs given by Straatman et al. (2014) may be under-estimated; for instance, a part of ZF may be star-forming and obscured by dust. \cite{Simpson17} argued that the strong Balmer features shown in ZF are not a unique signature of a post-starburst galaxy and are indeed frequently observed in infrared-luminous galaxies, and understanding high-redshift obscured starbursts will only be possible with multi-wavelength and high-resolution observations.

On the other hand, our AGN feedback model may under-estimate the abundance of the high-redshift massive QGs. In our simple AGN feedback model, the growth of a central AGN is triggered by a galaxy merger event, which is certainly over-simplified and not realistic. A more realistic model should also take other BH growth mechanisms into account, i.e., mergers might not be necessary to initialize the BH growth. In this sense, AGN feedback could kick in at even earlier epochs. In addition, the AGN feedback model implemented in Guo11 mainly refers to the `radio mode' feedback, which only operates on the cooling gas, yet does not affect the residual cold gas. However, when the AGN accretion rate is close to the Eddington limit, the `quasar mode' AGN feedback \citep{Croton06} could eject the residual cold gas from a galaxy in the form of a wind, with power at 5\--10\% of the accretion power \citep{Saez09,Dunn10}. This `quasar mode' AGN feedback is only partially represented by an enhanced feedback efficiency associated with starbursts (Croton et al. 2006) but is not fully incorporated into the current model. Therefore, in some galaxies with the moderate AGN accretion rates, both the `quasar mode' and `radio mode' feedback should operate to suppress star formation. As a result, AGN feedback should be more effective than what we assumed in this work. We thus expect to find more abundant massive QGs at $z\sim 3.7$ in a more realistic AGN feedback model.

In conclusion, the recently reported massive quiescent galaxy ZF at redshift $z=3.7$ is rare, but may appear naturally in the existing $\Lambda$CDM galaxy formation models.

\section*{Acknowledgments}
We thank Lan Wang, Ran Li and Chen Hu for useful discussions. LG acknowledges support from NSFC grants (nos. 11133003 and 11425312). LG and QG acknowledges support from Royal Society Newton advanced Fellowships. acknowledges support from the National Basic Research Program of China (program 973 under grant No. 2015CB857001) and NSFC under grant No. 11573030.

\bibliographystyle{mn2e}


\end{document}